\documentclass[preprint,pre,aps,showpacs]{revtex4}
\usepackage{graphicx}
\begin{document}

\title{Exploring the Lattice Gas Model for isoscaling}

\author{C. B. Das$^1$, and S. Das Gupta$^2$}

\affiliation{$^1$Physics Group, Birla Institute of Technology and Science, 
Pilani, Rajasthan, India 333 031} 

\affiliation{$^2$Physics Department, McGill University, 
Montr{\'e}al, Canada H3A 2T8}

\date{\today}

\begin{abstract}
Isotopic spin dependent lattice gas model is used to examine if it 
produces the isoscaling behaviour seen in intermediate energy heavy
ion collisions.  Qualitative features are reproduced but quantitative
agreement with experiments is lacking.

\end{abstract}


\maketitle
 
\section{Introduction}

A very interesting feature of heavy ion collisions at intermediate energy
is that provided the experimental set ups are identical,the ratio
of isotope yields from two different reactions, 1 and 2, $R_{21}(N,Z)=
Y_2(N,Z)/Y_1(N,Z)$ exhibits an exponential relationship as a function of
the isotope neutron number $N$ and proton number $Z$ 
\cite{Xu,Tsang1,Tsang2,Shetty1,Shetty2}:
\begin{eqnarray}
R_{21}(N,Z)=Y_2(N,Z)/Y_1(N,Z)=C\exp(\alpha N+\beta Z).
\end{eqnarray}
This is called isoscaling and $\alpha$ and $\beta$ are called isoscaling parameters.  
Various theoretical models have been examined
for at least an approximate emergence of this law.  A grand canonical model 
for multifragmentation
naturally leads to this law but it is also seen to emerge as an
approximate  formula from a canonical
model \cite{Chaudhuri1}, an expanding excited source model \cite{Tsang1} 
and even 
anti-symmetrised molecular dynamics calculations \cite{Ono}.  In this
note we investigate how well isoscaling is obeyed in the lattice gas model
(LGM). Part of what we do has already been looked at in a recent 
publication \cite{Su}
but both our perspectives and the lessons we derive from our study
are different.  Although highly schematic, LGM has been profitably utilised
for investigating phase transition properties and many features of
inclusive cross-sections.  One might expect that isoscaling is a general
feature, not depending upon finer details of models and hence should
be seen in the LGM multifragmentation model.  The advavantage of 
LGM is (1) it takes into account the interaction between nucleons,
both Coulomb and nuclear
exactly (although the nuclear interaction is very schematic); (2) it
takes into account excluded volume effects exactly and (3) decay
of hot nuclei need not be considered as the cluster algorithm
recognises only particle stable clusters.  Many other models can 
not handle the above three issues easily although there are other
many virtues in these other models.

%\section{Parameters in LGM}

The methods for calculating fragments in an isotopic spin dependent LGM have been described in 
many places (see, for example, \cite{Samaddar,Dasgupta1,Campi,Chomaz}).
The nearest neighbor bond between unlike particles $\epsilon_{np}$ is set at -5.33 MeV 
(to produce binding energy of 16 MeV per particle); and the bonds between like particles
$\epsilon_{pp}$ and $ \epsilon_{nn}$ are set at 0.  The justifications for 
these choices 
are explained in \cite{Dasgupta1}.  Coulomb energy between
protons is included.

These specifications are enough to allow one to calculate populations
of all fragments given the lattice size,
the mass and charge numbers of the dissociating system and the temperature.
But for later purposes we need to
estimate the value of symmetry energy
implied in the model.  The reason for this is the following.  The
parameter $\alpha$ (and also $\beta$ of eq.(1)) depends upon the value of
$N/Z$ of the the dissociating systems in reactions 1 and 2. Moreover, it also
depends upon the value of symmetry energy which adds a term $c_s(N-Z)^2/A$
to binding energy in the liquid-drop formula.  Indeed
it is a standard practice to use a relation \cite{Tsang1}
between $\alpha$ and the symmetry energy coefficient $c_s$, as:
\begin{eqnarray}
\alpha=\frac{4c_s}{T}[(Z/A)_1^2-(Z/A)_2^2].
\end{eqnarray}
It is therefore imperative to estimate a value of $c_s$ implied in the LGM.

\section{Estimation of $c_s$ in LGM}

We obtain ground state energies of a large number of nuclei by 
Monte-Carlo sampling at zero temperature.  For a given nucleus
with mass number $A$ and charge $Z$ this ground state
energy is denoted by $BE(LGM)$.  Let the value of the Coulomb energy 
which is also available from this calculation be $E_c$.  Since 
by choice the volume energy is $-16A$ MeV we can try to deduce the
value of surface energy coefficient $a_s$ and the symmetry energy
coefficient $c_s$ by setting
\begin{eqnarray}
BE(LGM)-E_c+16A=a_sA^{2/3}+c_s(N-Z)^2/A
\end{eqnarray}
Ideally the values of $a_s$ and $c_s$ should be the same for all
nuclei.  But because there is no good reason why $BE(LGM)$ should
obey this parametrisation exactly, values of $a_s$ and $c_s$
deduced from the above relation will change from nucleus to
nucleus.  We can now try to get the ``best'' values by minimising
the sum of the squares of the deviation. We chose
isotopes of some arbitrarily chosen $Z$'s.  The fit with the best 
values of $a_s$ and $c_s$ is shown in Fig.1.  

The fit in Fig.1 appears to be very good but some words of caution are needed.
Of the two constants we are trying to get, $a_s$ is 
by far the most important one; $a_sA^{2/3}$ dominates over $c_s(N-Z)^2/A$
which is a smaller perturbation.  But it is $c_s(N-Z)^2/A$ which is presumed
to be the deciding factor for isoscaling.  What is left after subtracting the
surface term can not be fitted by the parametrisation $c_s(N-Z)^2/A$
very accurately.  This is demonstrated in Fig.2.

What is reported here is similar to but not identical with the 
extraction of $a_s$ and $c_s$ in \cite{Samaddar}.  However there
the symmetry energy derived had a volume part (like here) but also
a surface part.  We have absorbed here all the effects of symmetry
energy using a volume term only in order to test how well LGM
calculations follow eq.(2) which is based on a volume symmetry energy. 

\section{Comparison with some data}

We will compare our calculations with two sets of data.  We first consider 
$^{112}$Sn+$^{112}$Sn (reaction 1) and $^{124}$Sn+$^{124}$Sn (reaction 2)
central collision data.  Experimental data are given in 
Fig.1 of \cite{Tsang2}.  Isoscaling is seen to be well satisfied
with a value of $\alpha$= 0.361.  In Fig.3. we
show calculated results for $R_{21}$ where this is the ratio of 
$\langle n_2(N,Z)
\rangle$ and $\langle n_1(N,Z)\rangle$; $\langle n_{N,Z} \rangle$
is the average multiplicity of the composite with $N$ neutrons and
$Z$ protons.  The dissociating systems are
taken to be $A$=168, $Z$=75 for reaction 1 and $A$=186, $Z$=75 for
reaction 2.  These are the recommended values \cite{Tsang2} after
allowing for losses like pre-equilibrium emissions etc.. 
The average multiplicity is calculated from 100,000 Monte-Carlo events. 
We try 10,000 switches between two events.  Metropolis algorithm
is used. The slopes of the ratios of the average multiplicities
should correspond to to the measured value of the slope of
experimental $Y_2(N,Z)/Y_1(N,Z)$ (Fig.1 of \cite{Tsang2}).

No basic calculation with the grand canonical model which computes
the value of $\alpha$ has been reported.  Canonical model
calculations are quite successful \cite{Das}.  SMM calculations before decay
of hot primaries show isoscaling quite well with $\alpha$=0.46;
but after decay of primaries isoscaling is not obeyed to the
same precision and the approximate $\alpha$ value changes slightly
to 0.44 \cite{Tsang2}.

Results from LGM with different lattice sizes (N) and at different 
temperatures (T) are shown in Fig.3.  
Isoscaling is obeyed very well though it is not as good as
in experimental data.  In LGM there is no
correction due to secondary decay.  The clusters calculated are
all particle stable.  Even though isoscaling is only approximately
obeyed we deduce an average value of $\alpha$.  It is about 0.20
for $8^3$ lattice and temperature 5 MeV (compared to 0.36 in experiment).
A notable feature of Fig.3 is that for $Z$=1, $N$ can be as high as 6
(Fig.3 shows results upto 5); $Z$=1 and $N$=6 as a stable composite  
happens with the proton in the central cube and 
six neutrons at the six faces.  The binding energy per particle for this
``nucleus'' is 5.33(6/7).  Of course in the real world such a nucleus
does not exist. 

We now examine how well eq.(2) is obeyed in LGM.  We can write
eq.(2) as $\alpha=f(c_s,T)g(1,2)$ where $g(1,2)$ is just a property
of the two dissociating systems and $f$ includes all the effects
of symmetry energy.  Examining the validity of $\alpha$ can be
ambiguous in our case as isoscaling is not equally good for
different $Z$'s.  Fig.3 shows that $Z$=2 obeys isoscaling quite
well.  Let us limit ourselves to $\alpha$ value for $Z$=2.
We consider $T$=4 MeV, lattice size $8^3$, keep $Z_1$ fixed
at 75, $A_1$ fixed at 168. For $Z_2$ fixed at 75 we vary $A_2$
from 168 to 186 and calculate $\alpha$ as $A_2$ is varied.  As a 
function of $(Z_2/A_2)^2$ the value of $\alpha$ is quite linear
(Fig.4) suggesting that the functional form of $g(1,2)$ of eq.(2) is  
accurate for LGM.  The same can not be said about $f(c_s,T)$.  If
we use eq.(2), $Z_1=Z_2=75$ and $A_1=168, A_2=186$ and $c_s=23.4$
MeV as obtained from the least square fit in section II, the
predicted value of $\alpha$ would be 0.86 (as contrasted with
$\alpha\approx 0.22$ as actually given by the LGM calculation).
A different point of view is sometimes taken.  One takes the
value of $\alpha$ and deduces the value of $c_s$ using eq.(2).
In our case from the value of $\alpha=0.22$ we would then be lead to
believe that the value of $c_s$ is way lower, about 6 MeV.  We
see no reason why a value of $c_s$ would so drastically change
from about 23 MeV for isolated nuclei to this low value at 
4/5 MeV temperature in an expanded volume in co-existence with
other hot nuclei.  It is specially hard to understand this in LGM.
The composites in LGM can not be squished or expanded.  At most,
some of the composites may be in excited states with 
slightly different number of bonds.  But such widely different value
for symmetry energy appears unlikely.  We remind the reader
that there are two parts to the calculations.  One is:
given the lattice size, the number of neutrons and protons,
the temperature and the bond strengths $\epsilon_{np},
\epsilon_{pp}$ and $\epsilon_{nn}$, calculation of thermodynamic properties
and many particle correlations at all levels.  Monte-Carlo
simulations solve this many-body problem of LGM correctly though
numerically.  Next comes the question: how does one define clusters,
given that the physics upto this point has been done correctly.
We follow the prescription, first formulated for LGM in \cite{Pan},
subsequently reformulated with the same result in \cite{Campi},
shown to be closely equivalent to the one derived in \cite{Coniglio}
and is now universally used.  A practical, reasonable but different
prescription is not known.
Given that the choice of the values of $\epsilon_{np},
\epsilon_{pp}$ and $\epsilon_{nn}$ is very restrictive, we have
no freedom to alter anything.  We find it easier to
believe that the function $f(c_s,T)$ as implied by eq.(2) is not correct
for LGM.

The other set of data we use is for $^{58}$Ni and $^{64}$Ni on $^9$Be.
We assume that the much larger nucleus Ni engulfs Be so that
for reaction 1 we have $A_1=67, Z_1=32$ and for reaction 2 we have
$A_2=73, Z_2=32$.  Experimental data can be found in 
\cite{Mocko1,Mocko2}.  The subset of data we use here can also be found 
in \cite{Chaudhuri1} (Figs.6 and 7).  Experimental results for Ni on
Be show a larger deviation from isoscaling compared to what is seen
experimentally for Sn on Sn.  Fig.5 shows the results of our 
calculation.  There are significant deviations from isoscaling.
The calculated value of $\alpha$ hovers around 0.22 (these averages
are always ambiguous since isoscaling deviations are more significant
here) whereas experimental values are around 0.6.  Thus, as in the 
case of Sn+Sn, theory underestimates the value of $\alpha$.

In the case of Ni on Be, experimental values of $Y_2(N,Z)/Y_1(N,Z)$
are available for a large range of $Z$ from 1 to 28.  Low and
moderate values of $Z$ have an effective $\alpha$
which is much smaller than those belonging to $Z\approx 22$.
(Data can be found in \cite{Chaudhuri1,Mocko1}.)
Such details do come out in LGM as well.  We demonstrate this 
in Fig.6.  This is highly satisfying.
Grand canonical model can not explain this difference although  
the canonical model model can and indeed fits the data very
well \cite{Chaudhuri1}.

\section{Summary}

Our first objective was to see if isoscaling is obtainable in
LGM.  It does appear that approximate isoscaling is obtained
in the model.  We confronted the calculations with two sets of
experimental data.  There are no free parameters in the
model.  The experimental values of $\alpha$ are larger than what 
the model predicts, by about a factor of 2.  But the model
does reproduce some remarkable features.  It did show a significant
increase in the value of $\alpha$ going from low $Z$ isotopes
to high $Z$ isotopes as seen in measurements.  It also gave
linearity of $\alpha$ with $\Delta(Z/A)^2$.  In view of the schematic 
nature of the model and obvious drawbacks, these successes are
quite pleasing.  Underestimation of the value of $\alpha$ remains a 
problem.  We have checked that reasonable variations of lattice size
or temperature will not correct this problem.  Moderate variations
in the values of the bond strengths do not provide enough corrections.

\section{Acknowledgement}
This research is supported by the Natural Science and Engineering Council
of Canada.  We thank G. Chaudhuri and A. Botvina for discussions and
Y. G. Ma for a communication.  C. B. Das thanks the physics dept.
of McGill University for hospitality during his visit in summer, 2008.

\newpage

\begin{figure}
\includegraphics[width=5.5in,height=6.5in,clip]{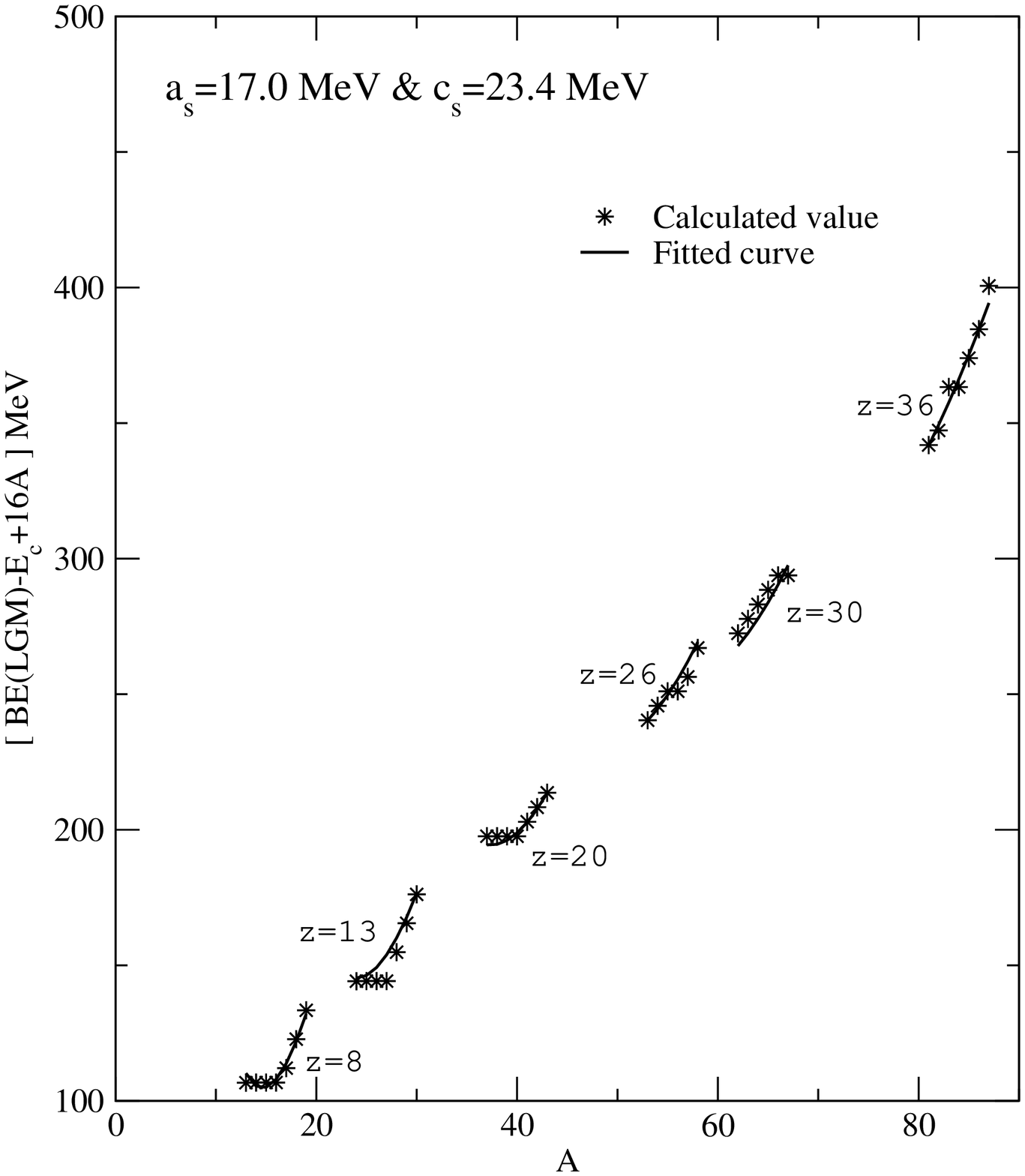}
\caption{For selected isotopes we show plots of $BE(LGM)-E_c+16A$ (stars)
and compare these with $a_sA^{2/3}+c_s(N-Z)^2/A$ where $a_s=17.0$
and $c_s=23.4$.  These values are chosen from least squares fit.  All
energies are in MeV.}
\end{figure}

\begin{figure}
\includegraphics[width=5.5in,height=6.5in,clip]{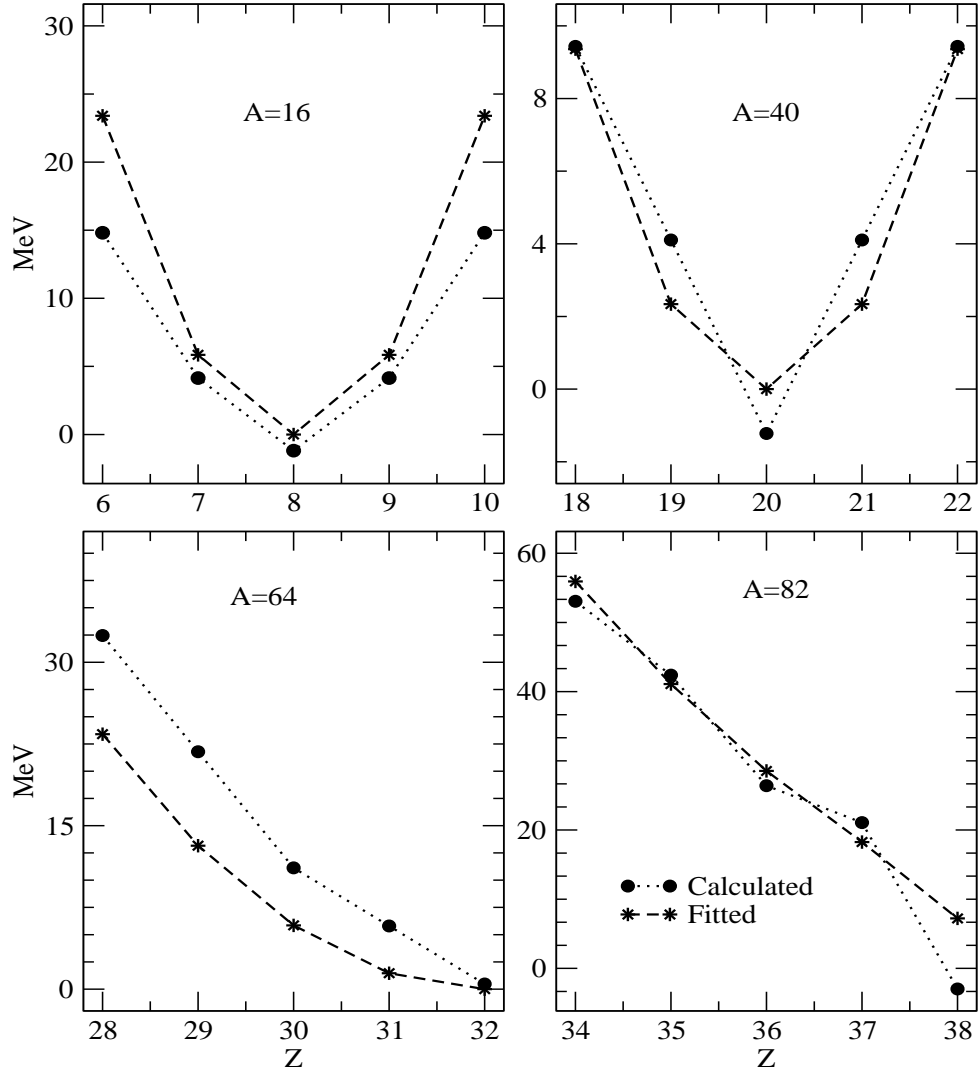}
\caption{ To test the accuracy of the parametrisation $c_s(N-Z)^2/A$
we plot $BE(LGM)-E_c+16A-17A^{2/3}$ and compare that with
$23.4*(N-Z)^2/A$.  All energies are in MeV.}
\end{figure}

\begin{figure}
\includegraphics[width=5.5in,height=6.5in,clip]{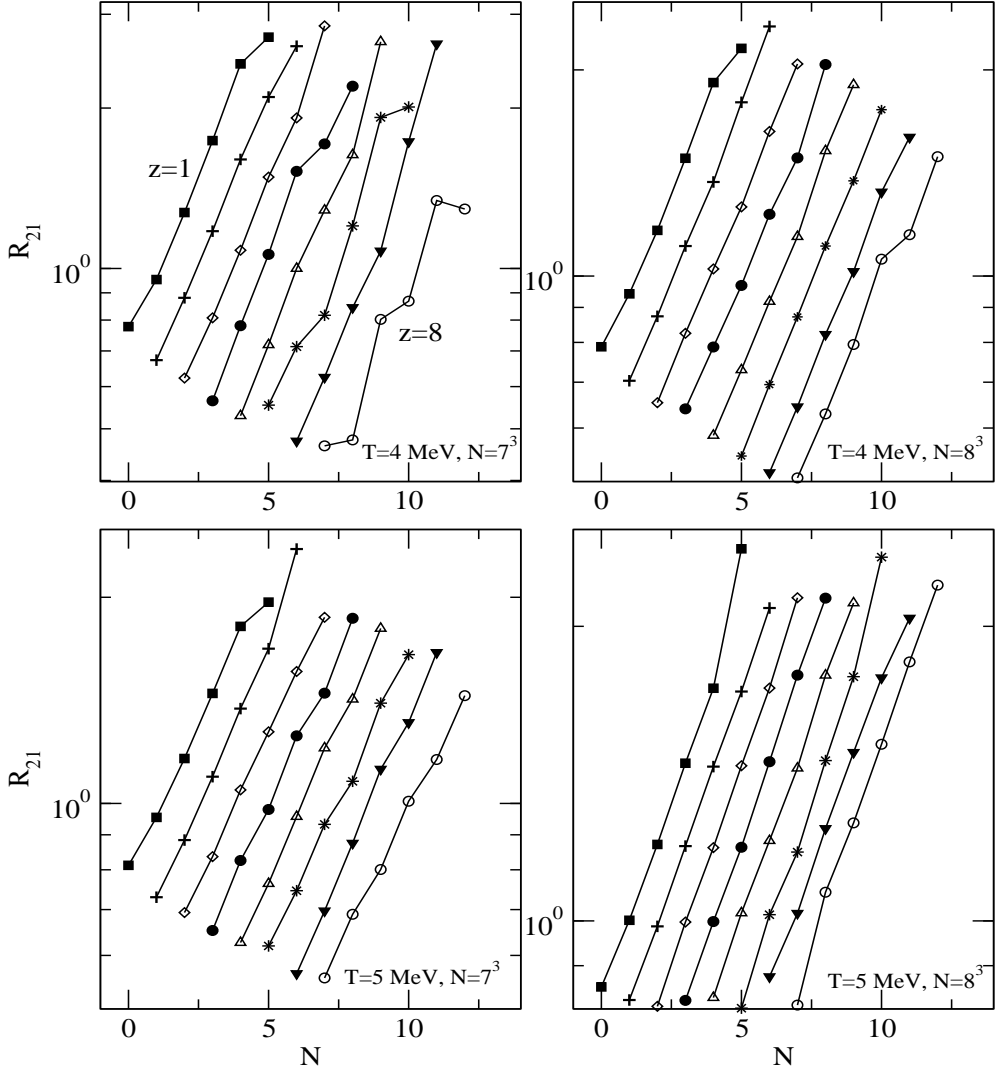}
\caption{These figures are calculated to compare with central collision
data for Sn on Sn reported in Fig.1 of \cite{Tsang2}.  For reaction
2, the dissociating system is set at $A=186, Z=75$.  For reaction 1
we take $A=168, Z=75$.  Isoscaling is approximately obeyed but
the deviations are not negligible.  The ``average'' value of
$\alpha$ is $\approx 0.2$.  Experimantally for this case isoscaling
is better obeyed and the value of $\alpha$ is $\approx 0.34$.}  
\end{figure}

\begin{figure}
\includegraphics[width=5.5in,height=6.5in,clip]{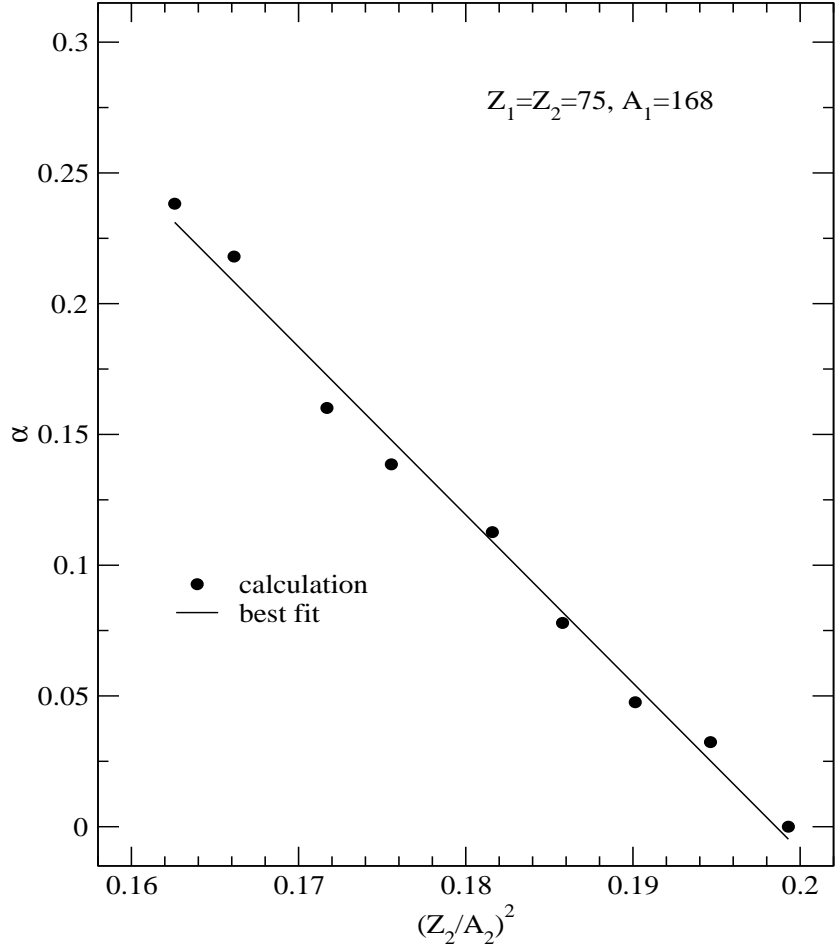}  
\caption{ For $Z_1=75$ and $A_1=168$ we vary $A_2$ from 168 to
186 for fixed $Z_2$=75.  The dots are from LGM calculation where
$\alpha$ is calculated for $Z=2$ as for this isotope isoscaling
is well obeyed (see Fig.3)).  The best linear fit is shown.  This comparison
tests the accuracy of the $[(Z_1/A_1)^2-(Z_2/A_2)^2]$ factor of
eq.(2).}
\end{figure}

\begin{figure}
\includegraphics[width=5.5in,height=6.5in,clip]{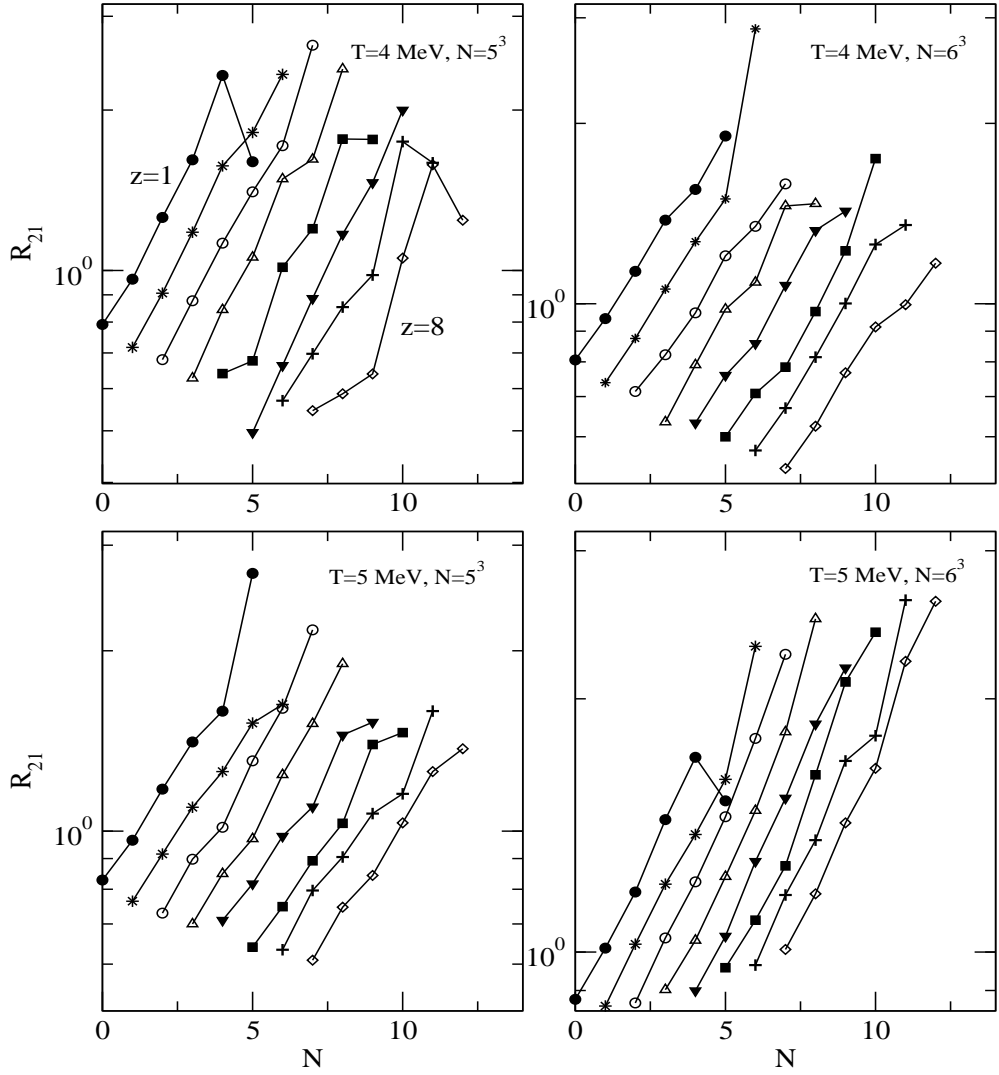}  
\caption{For reaction 2, the dissociating system is set at $A=73, Z=32$.
For reaction 1, the dissociating system is set at $A=67, Z=32$.
Experimental data for Ni on Be can be found in \cite{Mocko1} and
also in \cite{Chaudhuri1}, Figs 6 and 7.}
\end{figure}
  
\begin{figure}
\includegraphics[width=5.5in,height=6.5in,clip]{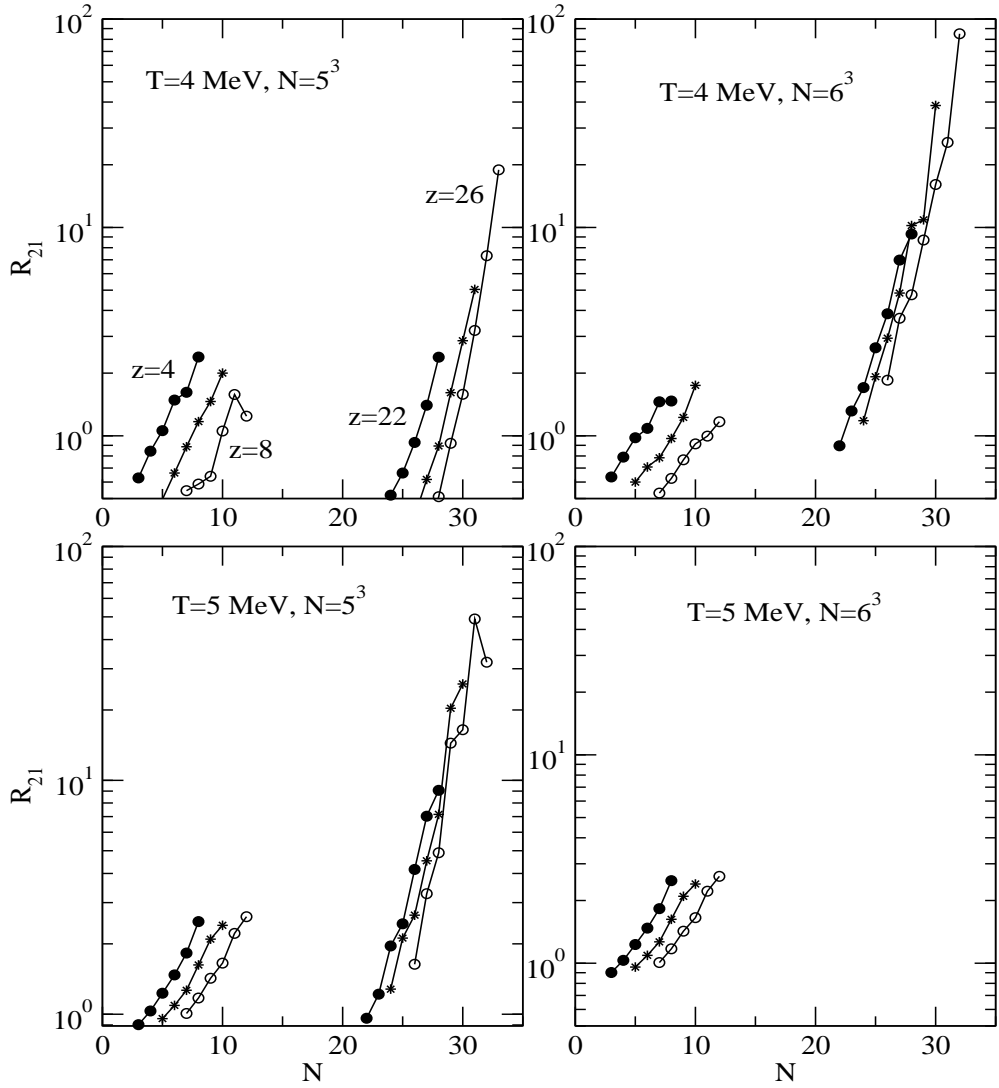}  
\caption{Same case as in Fig.5, but we also plot the ratios for much higher 
$Z$ isotopes.  Note the much higher value of the slopes for
large $Z$ nuclei.  Experimental data confirm this.  The high
$Z$ cases are not shown in the lowest right panel as there was not 
enough statistics to get a dependable ratio.}
\end{figure}

\end{document}